\documentclass[a4paper,11pt]{article}
\pdfoutput=1 % if your are submitting a pdflatex (i.e. if you have
\usepackage{jheppub} % for details on the use of the package, please
\usepackage[T1]{fontenc} % if needed

\usepackage[utf8]{inputenc}
\DeclareUnicodeCharacter{223C}{\ensuremath{\sim}}

\usepackage{braket}
\let\Ket\ket  % or \newcommand{\Ket}[1]{\ket{#1}}

\newcommand{\Hilbert}[2]{\mathcal{H}_{#1} \otimes \mathcal{H}_{#2}}

\newcommand{\relent}[2]{D\left(#1 \middle\| #2\right)}

\title{\boldmath Left-Right Relative Entropy}

\author{Mostafa Ghasemi}

\affiliation{Research Center for High Energy Physics, Sharif University of Technology,\\
	P.O.Box 11155-9161, Tehran, Iran \\
	School of Particles and Accelerators \& School of physics, Institute for Research in Fundamental Sciences (IPM) \\
	P.O.Box 19395-5531, Tehran, Iran }

\emailAdd{ghasemi.mg@ipm.ir}

\abstract{The concept of distinguishability lies at the heart of quantum information theory. We introduce \textit{left-right relative entropy} as a quantitative measure of distinguishability within the space of boundary states in two-dimensional conformal field theories (CFTs). By tracing over either the left- or right-moving modes, we derive a universal formula for arbitrary regularized boundary states defined on a circle. Remarkably, the resulting quantity reduces to a Kullback--Leibler divergence, where the associated probability distribution is determined entirely by the modular $\mathcal{S}$-matrix and the boundary data.
	For diagonal CFTs, we obtain exact expressions for the left-right relative entropy in terms of modular data, and extend the framework to define \textit{Sandwiched Left-Right R\'enyi relative entropies} and \textit{left-right fidelity}. Applying this formalism to the Ising model, tricritical Ising model, and $\widehat{su}(2)_k$ WZW model, we uncover a striking phenomenon: the left-right relative entropy between certain reduced boundary states vanishes even though the corresponding global boundary states are orthogonal. This observation motivates the introduction of \textit{relative entanglement sectors}, defined as equivalence classes of boundary states that are indistinguishable with respect to left-right relative entropy. These sectors transform as NIM-representations of global symmetries and exhibit level-dependent structures that mirror $\mathbb{Z}_2$ 't Hooft anomalies. Our findings establish an unexpected bridge between quantum information measures, boundary conformal symmetry, and quantum anomaly constraints.}

\begin{document} 
\maketitle
\flushbottom

\section{Introduction}
\label{sec:intro}

Quantum entanglement \cite{Horodecki:2009zz} is a defining characteristic of quantum systems that sets them apart from classical systems. It can be viewed as a measure of the non-local correlations between the components of these systems.
Quantum entanglement serves as a cornerstone for understanding various phenomena across condensed matter physics \cite{Kitaev:2005dm, Levin:2006LW, Dong:2008ft,Nayak:2008NA,	Vidal:2002rm, Amico:2007ag, Laflorencie:2015eck}, quantum field theory \cite{Calabrese:2009qy,Casini:2009sr}, and black hole physics \cite{Bombelli:1986, Srednicki:1993im}. Its applications not only enrich our theoretical frameworks but also challenge our understanding of reality, causality, and the fundamental nature of spacetime. 

One of the key measures of quantum entanglement is entanglement entropy(EE). For bipartite pure states, EE serves as an effective measure of quantum entanglement, defined as the von-Neumann entropy of the reduced density matrix $\rho_A$, $S_E = -\text{tr} (\rho_A \log \rho_A)$. 
This reduced density matrix is obtained by tracing out the degrees of freedom of the complementary region  $A^{c}$. 

Entanglement entropy is a fascinating concept, but it comes with its own set of challenges. Specifically, it exhibits ultraviolet divergences in relativistic field theories due to contributions from high-energy modes. Furthermore, its definition can be ambiguous in gauge theories \cite{Casini:2013rba}. In light of these difficulties, another information-theoretic quantity, relative entropy, can be considered. Although relative entropy is not a direct measure of entanglement, it is connected to various entanglement measures such as entanglement entropy, mutual information, and conditional entropy \cite{Vedral:2002zz}. Relative entropy is particularly valuable because it is proven to be ultraviolet finite, universal, and independent of gauge ambiguities\cite{Casini:2013rba}. These characteristics make it a robust measure for various applications in theoretical physics.

Relative entropy is a measure of distinguishability between quantum states and plays a crucial role in quantum information theory \cite{Vedral:2002zz}  and quantum statistical mechanics \cite{Wehrl:1978zz}. 
The relative entropy of two reduced  density matrices $\rho$ and $\sigma$ is defined as 
\begin{align}\label{DR-RE1}
	D(\rho\|\sigma) &= \text{tr}(\rho\log\rho)-\text{tr}(\rho\log\sigma).
\end{align}
The relative entropy value fall within the range of $[0,\infty]$, and $S(\rho\|\sigma)=0$ if and only if $\rho=\sigma$. 
This definition captures how distinguishable the two states are, with larger values indicating greater dissimilarity.  Relative entropy possesses several advantageous properties, such as monotonicity and positivity, making it a valuable tool in various areas of physics, including quantum field theory \cite{Faulkner:2016mzt, Balakrishnan:2017bjg,Casini:2016udt, Casini:2019qst}, conformal field theory \cite{Lashkari:2014yva, Lashkari:2015dia, Ruggiero:2016khg}, topological/conformal defect \cite{Ghasemi:2026hoa, Ghasemi:2026sij}, holography \cite{Blanco:2013joa, Wong:2013gua, Jafferis:2015del}, quantum gravity \cite{Casini:2008cr, Wall:2011hj, Longo:2018zib}, and  random states \cite{Kudler-Flam:2021rpr, Ghasemi:2024yzw}.

The one parameter generalization of relative entropy is known as sandwiched R\'enyi relative entropy \cite{		Tomamichel:2013bde,Wilde:2013bdg}
\begin{align}\label{REE01}
	D_{n}(\rho\|\sigma)= \frac{1}{n-1}\log\left[\text{tr}\left[\left( \sigma^{\frac{1-n}{2n}}\rho\sigma^{\frac{1-n}{2n}}\right) ^{n} \right]  \right] .
\end{align}
For certain values of $n$,  we can derive special quantities. Specifically, when  $n=2$ we are referring to collision relative entropy, and when $n=1/2$ we are referring to fidelity. Taking the limit $n\rightarrow 1$, we arrive at the relative entropy.
From a computational point of view, relative entropy can be computed through a replica trick \cite{Lashkari:2015dia}	
\begin{align}\label{REE2}
	D(\rho\|\sigma)= -\partial_{n}\,  \log \left(\frac{\text{tr}\left( \rho \sigma^{n-1}\right) }{\text{tr}\rho^{n}} \right) \Big|_{n \to 1}.
\end{align}
The author \cite{Lashkari:2014yva, Lashkari:2015dia} thoroughly investigated the concept of relative entropy in conformal field theory, emphasizing its practical application to arbitrary states within spatial subregions. This involves partitioning the Hilbert space and calculating the relative entropy for density matrices in excited states $\rho$ and $\sigma$ when they are reduced to the subsystem.

In this paper, we focus on the excited states of conformal field theories on a circle that exhibit conformal invariance and are influenced by the choice of boundary conditions. This class of states appears in  various contexts, including D-branes \cite{	Polchinski:1995mt,Gaberdiel:2003hm}, critical quantum quenches  \cite{Calabrese:2006rx, Calabrese:2007rg, Cardy:2014rqa,Calabrese:2016xau}, quantum impurity problems \cite{Affleck:1991tk,Affleck:1995ge} and the TQFT/CFT correspondence \cite{haldane-li,ludwig-qi}.
Conformal boundary states, denoted as $\Ket{B}$, are typically expressed as linear combinations of Ishibashi states, 
$| h_j \rangle\!\rangle$ \cite{ishibashi}. Generally, the state 
$\ket{B}$  is not normalizable and requires regularization.
One approach to regularize the norm is to implement Euclidean time evolution via 
$e^{-\epsilon H}$, where $H = \frac{2\pi}{\ell} ( L_0 + \bar{L}_0 - \frac{c}{12} )$. We then work with the regularized state defined as $\ket{\mathcal{B}}=\frac{e^{-\epsilon H}}{(\mathcal{Z}_{B})^{1/2}}\ket{B}$ instead \footnote{where $\epsilon$ is regarded as  UV parameter filter or quench parameter, it is identified by the inverse
	mass of the gapped theory, such that $\epsilon\approx m^{-1}$
	.}. This method ensures that the state remains well-defined while maintaining the necessary properties for further analysis.

Typically, conformal boundary states can be holomorphically factorized into their left and right-moving sectors. The entanglement entropy associated with this factorization is referred to as the left-right entanglement entropy (LREE). Previous studies have explored LREE in various contexts, including free bosons \cite{zayas,PandoZayas:2016cmz}, general CFTs \cite{Das:2015oha}, Chern-Simons theories \cite{Wen:2016snr}, chiral entanglement \cite{Lencses:2018paa}, reflected entropy \cite{Berthiere:2020ihq,Kusuki:2022bic}, and topological pseudo entropy \cite{Nishioka:2021cxe}.
In a similar vein, we define the relative entropy associated with left/right factorization as the \textit{left-right relative entropy} (LRRE). Using the boundary CFT formalism, we calculate LRRE for the states $\Ket{\mathcal{B}}$
in arbitrary CFTs defined on a circle of circumference $l$.
Our results yield a universal form for LRRE that is determined by the typical data of the CFT. It reduces to the Kullback-Leibler divergence with the  probability distribution  characterized by the modular $\mathcal{S}$-matrix and boundary conditions.

If we choose two Ishibashi states, $\rho_L^{(l)}=| h_l \rangle\!\rangle  \langle\!\langle h_l |$ and $\rho_L^{(k)}=| h_k \rangle\!\rangle  \langle\!\langle h_k |$, the LRRE becomes infinite due to the orthogonality of these states.
By selecting \textit{Cardy states} $\ket{\widetilde{l}}$ \cite{Cardy:1989ir, Cardy:2004hm}, which are  specific linear combinations of Ishibashi states $| h_a\rangle\!\rangle$ resulting  from imposing the modular invariance constraints
in diagonal rational CFTs, the LRRE can be expressed in terms of the modular $\mathcal{S}$-matrix
\begin{align}\label{0-bc}
	D_{\ket{\widetilde{h}_{l}},\ket{\widetilde{h}_{k}}} &=\sum_{j} |\mathcal{S}_{lj}|^2 \log  \frac{|\mathcal{S}_{lj}|^2}{|\mathcal{S}_{kj}|^2} .
\end{align}
and quantum fidelity becomes
\begin{equation}
	F(\rho_{L}^{(l)}\|\sigma_{L}^{(k)}) = \sum\limits_{j} |\mathcal{S}_{lj}| |\mathcal{S}_{kj}|.
\end{equation} 
We explicitly evaluate the left-right relative entropy (LRRE) corresponding to the Cardy states for the Ising model, the tricritical Ising model, and the $\widehat{su}(2)_{k}$
WZW model.  It is important to note that relative entropy is zero if and only if the two states are identical.

In the aforementioned models, the relative entropy between the left/right reduced density matrices of certain boundary states is zero, even though  these global states are orthogonal. From operational perspective, a chiral observer cannot distinguish boundary conditions that are related by center symmetry. This leads us to define the entanglement sector consisting of those boundary states with zero left/right relative entropy. We then identify the relative entanglement sectors for these models. Our findings suggest that the relative entanglement sector is intricately connected to the symmetry of the boundary states.

The paper is organized as follows. In Section \ref{sec:Left-Right} we present a detailed derivation of the left-right relative entropy (LREE), the sandwiched left-right R\'enyi relative entropies, and the left-right fidelity (LRF). We obtain universal formulas for the LREE and LRF for Ishibashi and Cardy states. In Section \ref{sec:Example}, we apply our formalism to the Ising model, the tricritical Ising model, and the $\widehat{su}(2)_k$ WZW model. We conclude with a discussion of implications and future directions in Section \ref{sec:Discu}. Additional details concerning the positivity of the LREE are provided in Appendix \ref{App:Positivity}, while its operational interpretation and remarks on the vanishing of the LRRE are given in Appendix \ref{App:Orthogonality}.

\section{Left-right Relative Entropy}
\label{sec:Left-Right}

We will consider rational conformal field theory (RCFT) whose Hilbert space takes the form
\begin{equation}
	\mathcal{H}=\bigoplus_{h,\bar{h}}n_{h,\bar{h}}\mathcal{V}_h\otimes \overline{\mathcal{V}}_{\bar{h}},
\end{equation}
where $\mathcal{V}_h$ and $\bar{\mathcal{V}}_{\bar h}$ are irreducible representations of the chiral and anti-chiral Virasoro algebras, and 
the non-negative integer $n_{h,\bar{h}}$ denotes the number of distinct primary fields with conformal weight $(h,\bar{h})$.
We focus on diagonal rational CFTs\footnote{%
	The framework naturally extends to irrational CFTs (Liouville theory, $c>1$; uncompactified free boson) with continuous spectra \cite{Teschner:2000md,Gaberdiel:2001zq}.%
}, characterized by $n_{h,\bar{h}} = \delta_{h,\bar{h}}$~\cite{yellowbook}. An arbitrary boundary state can be expressed as a linear combination of the Ishibashi states, $\ket{ B } =\sum _{a} \psi_{B}^a| h_a \rangle\!\rangle$ \footnote{These states satisfy the conformal boundary condition,  $L_{n} \ket{b} = \overline{L}_{-n}\ket{b}$. %where $L_n$ and $\overline{L}_{-n}$​	are the generators of the chiral and antichiral conformal transformations, respectively.
}.The boundary condition is encoded in the coefficients 
$\psi_B^{h_a}$. 

Ishibashi states can be expanded in terms of the orthonormal basis $\ket{h_a, N ; k}\otimes\ket{\overline{ h_a, N ; k}}$ in $\mathcal{V}_{h_a}\otimes \overline{\mathcal{V}}_{\bar{h}_a}$,  
\begin{align}
	| h_a \rangle\!\rangle = \sum_{N=0}^\infty  \sum_{k=1}^{d^{h_a}_N} \ket{h_a, N ; k} \otimes {\ket{\overline{ h_a,   N ;  k}}}.
\end{align}
The label $a$ denotes the primary field with weight $h_a$ and $d^{h_a}_N$ counts the degeneracy of descendants at each level $N$. 
The boundary state $\ket{B}$ is not normalizable, in the sense that it cannot be directly associated with a properly normalized density matrix, and one needs to regularize it \cite{zayas, Das:2015oha, Miyaji:2014mca}. Therefore, we utilize its regularized form
\footnote{We renormalize the Ishibashi states as $\langle\!\langle h_a \vert h_b \rangle\!\rangle = \delta_{ab} \mathcal{S}_{1a}$ \cite{Zuber:2000ia}. We can equally normalize the Ishibashi states as $\langle\!\langle h_a \vert h_b \rangle\!\rangle = \delta_{ab}$, which is the normalization employed in \cite{Wen:2016snr} and is equivalent to rescaling the coefficients as $\psi_{B}^a \to \psi_{B}^a / \sqrt{\mathcal{S}_{1a}}$.} ,
\begin{equation}\nonumber
	\ket{\mathcal{B}} = ( e^{-\epsilon H}/{{(\mathcal{N}_{B} )}^{1/2}}  )\ket{B},
\end{equation}
where $
\mathcal{N}_{\alpha}  
= \sum_{j} ({|\psi^{h_j}_{\alpha}|^2})\, \chi_{{h_j}}(e^{-8\pi \frac{\epsilon}{\ell}})$ represents the normalization factor and 
$
H = \frac{2\pi}{\ell} ( L_0 + \bar{L}_0 - \frac{c}{12} )
$ is the Hamiltonian. 
The character $\chi_{h_j}(e^{-8\pi\epsilon/\ell})$ of the highest-weight representation admits a universal expansion for unitary CFTs:
$
\chi_h(q) = \text{tr}_{\mathcal{V}_h}\!\left(q^{L_0-c/24}\right) = q^{-c/24}\sum_{N=0}^\infty d^h_N q^{h+N}$, where $q = e^{-8\pi\epsilon/\ell},
$ and $d^h_N$ counts states at level $N$ in the Verma module $\mathcal{V}_h$ \cite{yellowbook}.  
We choose two arbitrary regularized boundary states, $\ket{ B_{i} } =\sum _{a} \psi_{B_{i}}^a| h_a \rangle\!\rangle$, where $i=1,2$, and calculate LRRE for them.
The reduced density matrix associated with the regularized boundary state for the left-moving sector is derived by tracing out the right-moving sector,
\begin{align}\label{RDBS}
	&\rho_L^{(\alpha)} =  \frac{ 1 }{\mathcal{N}_{\alpha}}\sum\limits_{a, N, k} |\psi_{\alpha}^{h_a}|^2  e^{-8\pi \frac{\epsilon}{\ell}( h_a+ N -\frac{c}{24})}\ket{h_a, N, k }\bra{h_a, N, k},
	\nonumber\\&
	\sigma_L^{(\beta)}= \frac{ 1 }{\mathcal{N}_{\beta}}\sum\limits_{a, N, k} |\psi_{\beta}^{h_a}|^2  e^{-8\pi \frac{\epsilon}{\ell}( h_a+ N -\frac{c}{24})}\ket{h_a, N, k }\bra{h_a, N, k}.
\end{align}
To compute the LRRE, we employ the replica trick in the thermodynamic limit ($\ell/\epsilon \gg 1$). The diagonal structure of the reduced density matrices (\ref{RDBS}) enables exact evaluation of the trace of the $n^{\text{th}}$ power of them%
\footnote{This approach efficiently captures logarithmic terms, outperforming direct methods based on Eq. (\ref{DR-RE1}) and applying other techniques \cite{Kawahigashi:2004rf}.},
\begin{align}
	&\text{tr}_L  [\rho_L^{(\alpha)} ]^n= \ e^{\left[  {\frac{\pi \ell}{2\epsilon} \left(\frac{1}{n}-n\right)\frac{c}{24}} \right]} \  \frac{\mathcal{F}(n,\alpha)}{[\mathcal{F}(1,\alpha)]^n} ,
	\nonumber\\&
	\text{tr}_L[\rho_L^{(\alpha)}(\sigma_{L}^{(\beta)})^{(n-1)}] 
	\nonumber\\&
	\qquad \qquad =  \ e^{\left[  {\frac{\pi \ell}{2\epsilon} \left(\frac{1}{n}-n\right)\frac{c}{24}} \right]} \frac{\mathcal{G}(n,\alpha,\beta)}{[\mathcal{F}(1,\beta)]^{(n-1)}[\mathcal{F}(1,\alpha)]} \ .
\end{align}
Consequently,
\begin{align}
	G_{n}(\rho_{L}^{(\alpha)}\|\sigma_{L}^{(\beta)})&=\frac{\text{tr}\left( \rho_{L}^{(\alpha)} (\sigma_{L}^{(\beta)})^{n-1}\right) }{\text{tr}(\rho_{L}^{(\alpha)})^{n}} 
	\nonumber\\&
	=\frac{[\mathcal{F}(1,\alpha)]^{(n-1)}[\mathcal{G}(n,\alpha,\beta)]}{[\mathcal{F}(1,\beta)]^{(n-1)}[\mathcal{F}(n,\alpha)]},
\end{align}
where $\mathcal{F}(n,\alpha)$ and $\mathcal{G}(n,\alpha,\beta)$ are defined as follows:
\begin{align}
	&\mathcal{F}(n,\alpha) \ = \ {   \sum\limits_{a}  {|\psi_\alpha^{h_a}|^{2n}}      \,  \chi_{h_a}(e^{-8 \pi n\frac{\epsilon}{\ell}})    },
	\nonumber\\&
	\mathcal{G}(n,\alpha,\beta) \ = \ {   \sum\limits_{a}  {|\psi_\alpha^{h_a}|^{2}} {|\psi_\beta^{h_a}|^{2(n-1)}}     \,  \chi_{h_a}(e^{-8 \pi n\frac{\epsilon}{\ell}})   }.
\end{align}
Using the modular transformation property  of character $\chi_{h_a}(e^{-8 \pi n\frac{\epsilon}{\ell}})=\sum_{a'} \mathcal{S}_{aa'}\chi _{h_{a'}} (e^{-\frac{  \pi \ell}{2n\epsilon}})$, we obtain the following expression
in the thermodynamic limit $\ell/\epsilon \gg 1$,
\begin{align}
	G_{n}(\rho_{L}^{(\alpha)}\|\sigma_{L}^{(\beta)})
	=\frac{[\sum\limits_{a}  {|\psi_\alpha^{h_a}|^{2}}      \mathcal{S}_{1a} \,  ]^{(n-1)}}{[\sum\limits_{a}  {|\psi_\beta^{h_a}|^{2}}      \mathcal{S}_{1a} \,  ]^{(n-1)}}
	\frac{\sum\limits_{a}  {|\psi_\alpha^{h_a}|^{2}|\psi_\beta^{h_a}|^{2(n-1)}}      \mathcal{S}_{1a} \,  }{\sum\limits_{a}  {|\psi_\alpha^{h_a}|^{2n}}      \mathcal{S}_{1a} \,}.  
\end{align}
We obtain the relative entropy by taking the derivative with respect to $n$ and the limit of $n\rightarrow 1$,
\begin{align}\label{U-TRE}
	D(\rho_{L}^{(\alpha)}\|\sigma_{L}^{(\beta)})& = -\, \partial_n \log G_{n}(\rho_{L}^{(\alpha)}\|\sigma_{L}^{(\beta)})\Big|_{n \to 1} \nonumber\\&
	= \frac{\sum\limits_a \mathcal{S}_{1a} |\psi_{\alpha}^{h_a}|^2 \log(|\psi_{\alpha}^{h_a}|^2 )}{ \sum\limits_a \mathcal{S}_{1a  }|\psi_{\alpha}^{h_a}|^2 } -\log  \sum\limits_a \mathcal{S}_{1a} |\psi_{\alpha}^{h_a}|^2
	\nonumber\\&
	- \frac{\sum\limits_a \mathcal{S}_{1a} |\psi_{\alpha}^{h_a}|^2 \log(|\psi_{\beta}^{h_a}|^2 )}{ \sum\limits_a \mathcal{S}_{1a  }|\psi_{\alpha}^{h_a}|^2 } + \log  \sum\limits_a \mathcal{S}_{1a} |\psi_{\beta}^{h_a}|^2. 
\end{align}
This is the general formula for the left-right relative entropy and is free from any cut-off dependent divergence for any choice of $\psi_i^{h_a}$, as we naturally expect from relative entropy. By selecting specific boundary states, 
we demonstrate that the left-right relative entropy can be fully expressed in terms of the theory's modular  $\mathcal{S}$ matrix. 
It provides a key measure for quantifying the dissimilarity between two boundary states.

We define a probability distribution  $p^{(\alpha)}_{ a}$ which is characterized by the modular $\mathcal{S}$-matrix and boundary conditions as
\begin{align}\label{PROB-DIST}
	p^{(\alpha)}_{ a} &\equiv \frac{\mathcal{S}_{a1}}{\sum_{i}\,\mathcal{S}_{a1} |\psi_{\alpha}^{h_a}|^2}|\psi_{\alpha}^{h_a}|^2 \ , 
	\nonumber\\
	p^{(\beta)}_{ a} &\equiv \frac{\mathcal{S}_{a1}}{\sum_{i}\,\mathcal{S}_{a1} |\psi_{\beta}^{h_a}|^2}|\psi_{\beta}^{h_a}|^2  \ .
\end{align}
LRREE can now be written as
\begin{align}\label{U-TRE2}
	D(\rho_{L}^{(\alpha)}\|\sigma_{L}^{(\beta)})& =
	\sum\limits_a p^{(\alpha)}_{ a} \log(\frac{p^{(\alpha)}_{ a}}{p^{(\beta)}_{ a}} )
\end{align}
With this definition, the left-right relative entropy reduces to the Kullback-Leibler divergence \cite{KullbackLeibler}, which is non-negative quantity.

\subsection{Sandwiched Left-Right R\'enyi relative entropies and Left-Right Fidelity}
\label{subsec:Sandwiched R\'enyi}

In a similar manner, we can derive an expression for the sandwiched Renyi relative entropy:
\begin{align}\label{SNRE-1}
	&D_{n}(\rho_{L}^{(\alpha)}\|\sigma_{L}^{(\beta)})
	\nonumber\\&
	=\frac{1}{n-1}\log\left[ \frac{\sum\limits_{a}  {|\psi_\alpha^{h_a}|^{2n}|\psi_\beta^{h_a}|^{2(1-n)}}\mathcal{S}_{1a} \,}{[\sum\limits_{a}  {|\psi_\alpha^{h_a}|^{2}}      \mathcal{S}_{1a} \,  ]^{n}[\sum\limits_{a}  {|\psi_\beta^{h_a}|^{2}}      \mathcal{S}_{1a} \,  ]^{(1-n)}}\right] 	\nonumber\\&
	=\frac{1}{n-1}\log\sum\limits_{a}(p^{(\alpha)}_{ a})^{n}\, (p^{(\beta)}_{ a})^{(1-n)} 
	.  
\end{align}
The expression (\ref{SNRE-1}) for the special value of $n=\frac{1}{2}$ is
related to the fidelity, which is a natural generalization of the notion of pure states overlap, $\arrowvert \langle\phi|\psi\rangle\arrowvert$. Quantum fidelity can be used as a proper tool in characterizing quantum phase  transition \cite{Shi-Gu:2008zq}. Left-right quantum fidelity was obtained as 	
\begin{align}\label{FID-1}
	&F(\rho_{L}^{(\alpha)}\|\sigma_{L}^{(\beta)})
	=\frac{\sum\limits_{a}  {|\psi_\alpha^{h_a}||\psi_\beta^{h_a}|}\mathcal{S}_{1a} \,}{[\sum\limits_{a}  {|\psi_\alpha^{h_a}|^{2}}      \mathcal{S}_{1a} \,  ]^{\frac{1}{2}}[\sum\limits_{a}  {|\psi_\beta^{h_a}|^{2}}      \mathcal{S}_{1a} \,  ]^{\frac{1}{2}}}	\nonumber\\&
	=\sum\limits_{a} \sqrt{p^{(\alpha)}_{ a}}\,\sqrt{p^{(\beta)}_{ a}}
	.  
\end{align}
It reduces to the fidelity between two probability distributions defined in Eq. (\ref{PROB-DIST}).

\subsection{Ishibashi and Cardy States}
\label{subsec:Ishibashi and Cardy}

If we specifically choose the boundary states $\ket{B_{i}}$ to be two Ishibashi states, $\rho_L^{(j)}=| h_j \rangle\!\rangle  \langle\!\langle h_j |$ and $\sigma_L^{(k)}=| h_k \rangle\!\rangle  \langle\!\langle h_k |$
-- equation \eqref{U-TRE}  with $\psi_{\alpha}^{h_a}=\delta_{aj}$ and $\psi_{\beta}^{h_a}=\delta_{ak}$-- and using the above definition,
the associated left-right relative entropy becomes
\begin{align}\label{TRE0}
	D\left(\rho_L^{(j)}\|\sigma_L^{(k)}\right)&=\infty.
\end{align}
It is infinite due to the orthogonality of two Ishibashi states.
Using the relation (\ref{FID-1}), the left-right quantum fidelity becomes
\begin{equation}
	F(\rho_{L}^{(j)}\|\sigma_{L}^{(k)}) = \frac{\sum\limits_{a}  {\delta_{aj}\delta_{ak}}\mathcal{S}_{1a} \,}{[\mathcal{S}_{1j} \,  ]^{\frac{1}{2}}[\mathcal{S}_{1k} \,  ]^{\frac{1}{2}}}=
	\begin{cases}
		1 & \text{if $j = k$}  \\
		0 & \text{otherwise}
	\end{cases}
\end{equation}

Cardy states are specific linear combinations of Ishibashi states \cite{Cardy:1989ir, Cardy:2004hm, Gaberdiel:2003hm},
\begin{align}\label{CARD-state}
	\ket{B_{1}} \equiv \ket{{\widetilde{h}_l}} &= \sum_j \frac{\mathcal{S}_{lj}}{\sqrt{\mathcal{S}_{1j}}} | h_j \rangle\!\rangle,\nonumber\\
	\ket{B_{2}} \equiv \ket{{\widetilde{h}_k}} &= \sum_j \frac{\mathcal{S}_{kj}}{\sqrt{\mathcal{S}_{1j}}} | h_j \rangle\!\rangle.
\end{align}
The boundary coefficients are defined as $\psi_{l}^j = \mathcal{S}_{lj}/\sqrt{ \mathcal{S}_{1j} }$ and $\psi_{k}^j = \mathcal{S}_{kj}/\sqrt{\mathcal{S}_{1j} }$. With the real, symmetric, and unitary properties of $\mathcal{S}$, the result \eqref{U-TRE} can now be completely expressed in terms of the modular $\mathcal{S}$-matrix. Thus, the LRRE for Cardy states in diagonal CFTs is:
\begin{align}\label{TRE-CARD}
	D_{\ket{\widetilde{h}_{l}},\ket{\widetilde{h}_{k}}} &=\sum_{j} |\mathcal{S}_{lj}|^2 \log  \frac{|\mathcal{S}_{lj}|^2}{|\mathcal{S}_{kj}|^2}. 
\end{align}
The equations (\ref{U-TRE}) and (\ref{TRE-CARD}) demonstrate that the terms are finite and do not depend on the scales (UV cutoff) involved in the system, consistent with our understanding of relative entropy. The quantum fidelity is then: 
\begin{equation}
	F(\rho_{L}^{(l)}\|\sigma_{L}^{(k)}) = \sum\limits_{j} |\mathcal{S}_{lj}| |\mathcal{S}_{kj}|.
\end{equation} 
In the context of equation (\ref{TRE-CARD}), it's worth noting that the denominator $\mathcal{S}_{kj}^2$ can be zero, leading to the possibility of the left-right relative entropy becoming infinite. It also is positive \footnote{For the proof of the positivity of Eq.(\ref{TRE-CARD}) See the Appendix \ref{App:Positivity}.}. Moreover, as we will demonstrate in the upcoming examples, it's possible for the left-right relative entropy between the left/right reduced density matrices of two boundary states to be zero, despite their orthogonality as global states \footnote{As we can see, the relative entropy between two density matrices is zero if and only if the density matrices are equal. However, this does not necessarily mean that the global states (which for example, purify the density matrices) are equal as well. See the Appendix \ref{App:Orthogonality} for a more detailed explanation, along  with an example.}.

\section{Example}
\label{sec:Example}

In this section, we will illustrate specific examples to demonstrate how to calculate the left-right relative entropy. Equation \eqref{TRE-CARD} demonstrates that having the modular $\mathcal{S}$ matrix enables the direct computation of the LRRE for the Cardy states \eqref{CARD-state}. We will explicitly compute the LRRE for three models: the Ising model, the tricritical Ising model, and the  $\widehat{su}(2)_{k}$  WZW model.

Ising model: The critical Ising CFT ($c = 1/2$) is equivalent to a $\mathbb{Z}_2$ orbifold of a free massless Majorana fermion CFT via 2D bosonization.
It contains  three primary operators : the identity $\mathbb{I}$, the thermal operator $\varepsilon$, and the spin field $\sigma$
with conformal weights of $
0 , \ 1/2 , \ 1/16
$, respectively. 
The modular $\mathcal{S}$ matrix for the Ising model can be found in \cite{yellowbook}.
When the boundary state $\ket{\widetilde 0} $ is taken as the reference state, the left-right relative entropy for the Ising model is determined by the following expression:
\renewcommand{\arraystretch}{1.5}
\begin{center}
	\begin{tabular}{  c l }
		\hline
		\ \ \ \ Boundary condition\ \ \ \                                        & \ \ \ \ \ \ \ \ \ \ LRRE\ \ \ \ \ \ \ \ \ \ \   LRF \ \ \ \\ \hline\hline
		$\ket{\widetilde{ \frac{1}{2}}}$, \quad $\ket{\widetilde 0}$ &\ \ \ \ \ \ \ \ \ \ \ \ \ $ 0\ \ \ \  $ \quad  \ \ \ \ \ \ \ $ 1\ \ \ \ $ \\
		$\ket{\widetilde{ \frac{1}{16}}}$, \quad  $\ket{\widetilde 0}$           & \ \ \ \ \ \ \ \ \ \ \ \ \ $ \log 2 \ $    \quad  \ \ \ $ \frac{\sqrt{2}}{2}\ \ \ \  $    
		\\ \hline
	\end{tabular}
\end{center} \null 
We introduce the concept of the relative entanglement sector (RES), defined as the set of boundary states with zero left-right relative entropy. In the case of the critical Ising model, the relative entanglement sectors are $E_{1}\bigoplus E_{2}$, where $E_{1}$ consists of  $\{\ket{\widetilde 0},\ket{\widetilde{ \frac{1}{2}}}\}$, and $E_{2}$ consists of $\{\ket{\widetilde{\frac{1}{16}}}\}$.

From the lattice perspective, the Cardy states
$\ket{\widetilde 0}$, $\ket{\widetilde{ \frac{1}{2}}}$, and
$\ket{\widetilde{ \frac{1}{16}}}$
are  represented by \( \lvert \uparrow \rangle \), \( \lvert \downarrow \rangle \), and \( \lvert f \rangle \), respectively. The state \( \lvert \uparrow \rangle \) represents a boundary condition where the boundary spin is fixed in the up direction, while the state \( \lvert \downarrow \rangle \) corresponds to a spin fixed in the down direction. In contrast, the state \( \lvert f \rangle \) denotes a free boundary condition, allowing the boundary spin to fluctuate freely.

The Ising model exhibits a non-anomalous \( \mathbb{Z}_2 \) symmetry, generated by \( \eta \). This symmetry transforms the spin field \( \sigma \) and its descendants as \( \sigma \to -\sigma \), while leaving the identity operator  $\mathbb{I}$ and the energy operator \( \epsilon \), along with their descendants, unaffected.
The action of the \( \mathbb{Z}_2 \) symmetry on the Cardy states is defined as follows:
\begin{equation}
	\hat{\eta} \lvert \uparrow \rangle = \lvert \downarrow \rangle, \quad \hat{\eta} \lvert \downarrow \rangle = \lvert \uparrow \rangle, \quad \hat{\eta} \lvert f \rangle = \lvert f \rangle.
\end{equation} 
This definition shows that the operator \( \hat{\eta} \) exchanges the fixed boundary states \( \lvert \uparrow \rangle \) and \( \lvert \downarrow \rangle \), while leaving the free boundary state \( \lvert f \rangle \) invariant. Specifically, $E_{1}$  form a 2-dimensional NIM-rep \footnote{The NIM-reps are matrix representations with non-negative integer entries \cite{Fuchs:2001qc}.%, referred to as \textit{non-negative integer matrix representations	}
}, the smallest nontrivial permutation representation of \( \mathbb{Z}_2 \), while  $ E_{2}$ corresponds to the trivial representation  \cite{Choi:2023xjw}. %It indicates  that \( \lvert f \rangle \) is a \( \mathbb{Z}_2 \)-symmetric boundary, whereas \( \lvert \uparrow \rangle \) and \( \lvert \downarrow \rangle \) are not 

Tricritical Ising model:
The tricritical Ising model is the Virasoro minimal model $\mathcal{M}(5,4)$ with $c=7/10$. This theory contains six primary operators %-- an identity, %($\mathbbm{I}$), three energy operators %($\varepsilon, \ \varepsilon ' , \ \varepsilon ''$) and two Ising spins% ($\sigma, \ \sigma '$) --
with conformal dimensions
$
0  , \  {1}/{10}   , \ {3 / 5}   , \ {3/ 2}   , \ {3 / 80}   
$ and $ {7 / 16} $.
%The characters $\chi_{r,s}$ corresponding to the primaries $h_{r,s}$ are 
%$
%\chi_{1,1}   , \ \chi_{3,3}  , \ \chi_{2,3}  , \ \chi_{1,3}   , \ \chi_{2,2}   , \ \chi_{1,2} 
%$. 
The results for the LRRE, for various boundary states, are listed in the following table. (Here, $s_1=\frac{1}{\sqrt{5}}\sin\frac{2\pi}{5}$ and $s_2=\frac{1}{\sqrt{5}}\sin\frac{4\pi}{5}$. These appear as entries in the modular $\mathcal{S}$ matrix \cite{yellowbook,Castro:2011zq}).

\renewcommand{\arraystretch}{1.5} 
\begin{center}
	\begin{tabular}{  c l } 
		\hline
		Boundary condition\ \ & \ \ \ \ \ \ \ \ \ \ \ \ \ \ \ \ \ \ \ \ \ \ \ \ \ \ \ \ \ \ \ \ \ \   \ \ \ \ LRRE \ \   \\ 
		\hline
		\hline
		$\ket{\widetilde{\frac{3}{2}}}$, $\ket{\widetilde{0}}$ &\ \ \ \ \ \ \ \ \ \ \ \ \ \ \ \ \ \ \ \ \ \ \ \ \ \ \ \ \ \ \ \ \ \ \ \ \ \ \ \ $ 0  $\\
		${\ket{\widetilde{\frac{1}{10}}}}$, $\ket{\widetilde{0}}$ &\ \ \ \ \ \ \ \ \ \ \ \ \ \ \ \ \ \ \ \ \ \ \ \ \ \ \ \ \ \ \ \ \ \ \ \ \ \ \ \ $ 8({\left(\mathit{s}_1^2- \mathit{s}_2^2\right) \left( \log \mathit{s}_1\right)+\left(\mathit{s}_2^2- \mathit{s}_1^2\right) \left(\log\mathit{s}_2  \right)})  $\\
		${\ket{\widetilde{\frac{3}{5}}}}$, $\ket{\widetilde{0}}$ &\ \ \ \ \ \ \ \ \ \ \ \ \ \ \ \ \ \ \ \ \ \ \ \ \ \ \ \ \ \ \ \ \ \ \ \ \ \ \ \ $ 8({\left(\mathit{s}_1^2- \mathit{s}_2^2\right) \left( \log \mathit{s}_1\right)+\left(\mathit{s}_2^2- \mathit{s}_1^2\right) \left(\log\mathit{s}_2  \right)})  $\\
		${\ket{\widetilde{\frac{7}{16}}}}$, $\ket{\widetilde{0}}$ & \ \ \ \ \ \ \ \ \ \ \ \ \ \ \ \ \ \ \ \ \ \ \ \ \ \ \ \ \ \ \ \ \ \ \ \ \ \ \ \ $8\mathit{s}_1^2\log\left(\frac{\mathit{s}_1}{\mathit{s}_2}\right) +4\left(\mathit{s}_2^2- \mathit{s}_1^2\right)\log2 $
		\\
		${\ket{\widetilde{\frac{3}{80}}}}$, $\ket{\widetilde{0}}$ & \ \ \ \ \ \ \ \ \ \ \ \ \ \ \ \ \ \ \ \ \ \ \ \ \ \ \ \ \ \ \ \ \ \ \ \ \ \ \ \ $ 4\left(\mathit{s}_1^2+ \mathit{s}_2^2\right)\log2 $
		\\
		\hline
	\end{tabular}
\end{center}

In this model, the non-trivial relative entanglement sectors are $\{\ket{\widetilde 0},\ket{\widetilde{ \frac{3}{2}}}\}$ and $\{{\ket{\widetilde{\frac{3}{5}}}},\ket{\widetilde{\frac{1}{10}}}\}$.

\def\suck{\widehat{su}(2)_{k}}
The $\widehat{su}(2)_{k}$ WZW model:
The $\suck$ WZW model has a  central charge of $c=3k/(k+2)$ and Cardy boundary states are labeled by spins \( j \), where \( j = 0, \frac{1}{2}, 1, \ldots, \tfrac{k}{2}\). 
The model exhibits a $\mathbb{Z}_2$ center symmetry, which acts on boundary states as:
$|\widetilde{j}\rangle \mapsto \left| \tfrac{k}{2} - \widetilde{j}\right\rangle$, corresponding to a nontrivial automorphism of the fusion ring. Using the theory's modular $\mathcal{S}$ matrix \cite{Fendley:2006gr}, the LRRE from equation \eqref{TRE-CARD} is straightforward. For the diagonal mass matrix $\mathcal{M}=\mathbb{I}$, we find
\begin{equation}\label{LRRE-WZW}
	D_{j,j'} = \sum_{l=0}^{k/2} p_j(l) \log \left( \frac{p_j(l)}{p_{j'}(l)} \right),
\end{equation}
where $p_j(l) = \frac{2}{k+2} \sin^2 \theta_{jl}$ and $\theta_{jl} \equiv \frac{\pi (2j+1)(2l+1)}{k+2}$. 
The relative entanglement sectors of $\widehat{su}(2)_{k}$ WZW  models consist of states with zero LRRE. This condition holds if and only if:  $j = j' \quad \text{or} \quad j = \tfrac{k}{2} - j'$, reflecting $\mathbb{Z}_2$ symmetry equivalences among boundary states.

For RES sectors defined by \([j] = \{j, \tfrac{k}{2} - j\}\), even \(k\) admits a unique fixed point at \(j = k/4\), yielding a non-anomalous singlet, while all other sectors form \(\mathbb{Z}_2\) doublets. For odd \(k\), all sectors are doublets with no fixed point. The structure of relative entanglement sectors reproduces the anomaly-sensitive fixed-point structure of the \(\mathbb{Z}_2\) center symmetry, e.g, 't Hooft anomaly in boundary states at odd and even levels, respectively \cite{Numasawa:2017crf}.

For  $\widehat{su}(2)_{k}$, the RES sectors coincide with the orbits of the $\mathbb{Z}_2$ symmetry $j \leftrightarrow k/2 - j$. The fixed-point structure of these orbits (present for even $k$, absent for odd $k$) matches the known anomaly classification of the $\mathbb{Z}_2$ symmetry in this model. This suggests that LRRE may encode anomaly data, though a direct derivation is left for future work.

For example, let's set $k=2$. Boundary states in the  $\widehat{su}(2)_{2}$ WZW model correspond to primary fields with spins $
j = 0, \frac{1}{2}, 1$.
These boundary states are organized under a \(\mathbb{Z}_2\) symmetry as follows:
A symmetric pair of boundary states
$
\{|\widetilde{0}\rangle, |\widetilde{1}\rangle\},
$
and a single \(\mathbb{Z}_2\) invariant boundary state
$
|\widetilde{\frac{1}{2}}\rangle.
$
\begin{center}
	\begin{tabular}{  c l }
		\hline
		\ \ \ \ Boundary condition\ \ \ \                                        & \ \ \ \ \ \ \ \ \ \ LRRE \ \  \ \ \ \ \ \   LRF \ \ \ \\ \hline\hline
		$\ket{\widetilde{ 1}}$, \quad $\ket{\widetilde 0}$ &\ \ \ \ \ \ \ \ \ \ \ \ \ $ 0\ \ \ \  $ \quad  \ \ \ \ \ \ \ $ 1\ \ \ \  $ \\
		$\ket{\widetilde{ \frac{1}{2}}}$, \quad  $\ket{\widetilde 0}$           & \ \ \ \ \ \ \ \ \ \ \ \ \ $  \log 2 \ $    \quad  \ \ \ $ \frac{\sqrt{2}}{2}\ \ \ \ $    
		\\ \hline
	\end{tabular}
\end{center} \null 
The relative entanglement sectors are \( \{|\widetilde{0}\rangle, |\widetilde{1}\rangle\} \) and \( \{|\widetilde{\tfrac{1}{2}}\rangle\} \). This structure realizes the Tambara-Yamagami (TY) fusion category \( \mathrm{TY}_{\mathbb{Z}_2}^- \) \cite{Choi:2023xjw}. Similar computations of LRRE can be carried out for higher-rank WZW models or coset constructions.

Regarding the vanishing of left-right relative entropy between two reduced density matrices, this phenomenon occurs exclusively when their global boundary states are related by global symmetries (e.g., the $\mathbb{Z}_2$ symmetry in $\widehat{su}(2)_k$ WZW models).
Operational meaning: A chiral observer cannot distinguish boundary conditions that are related by center symmetry.

\section{Discussion}
\label{sec:Discu}

In this work, we have introduced the concept of  left-right relative entropy, as a measure of distinguishability in the space of boundary states. It is reduced to the Kullback-Leibler divergence with the  probability distribution  characterized by the modular $\mathcal{S}$ matrix and boundary conditions. We also derived a universal formula for the LRRE, the Sandwiched Left-Right R\'enyi relative entropies, and Left-Right Fidelity for boundary states in generic CFTs on a circle. Notably, these universal formula are free of UV divergences for any choice of boundary states. 
Furthermore, in the case of diagonal CFTs with a specific choice of boundary state, we demonstrated that the LRRE can be completely expressed in terms of the theory's modular $\mathcal{S}$ matrix. Moreover, we discovered that the left-right relative entropy between left/right reduced density matrices associated with certain boundary states can be zero, despite them not being identical as global states. These novel findings led us to define the concept of the relative entanglement sector, representing the set of boundary states characterized by zero left-right relative entropy. Our results suggest a profound connection between relative entanglement sectors and the symmetry properties of boundary states, indicating that these sectors transform as  NIM-reps of global symmetries in the underlying theory. Moreover, their level-dependent pattern analogous to \( \mathbb{Z}_2 \) 't Hooft anomalies in the boundary states.

The implications of zero left-right relative entropy for boundary conformal field theories are significant and multifaceted. When the left-right relative entropy between two reduced density matrices associated with global boundary states is zero, it indicates that these states are indistinguishable in terms of their physical properties, i.e., certain boundary conditions can lead to equivalent physical descriptions. This equivalence simplifies the analysis of entanglement properties and correlations within the system.
This suggests that they may exhibit similar behaviors under various transformations, which is crucial for understanding the symmetry and dynamics of the system.

For instance, in the Ising model, the boundary states in the relative entanglement sectors transform into one another under the action of the \( \mathbb{Z}_2 \) symmetry. In other words, the relative entanglement sector formed by the two fixed boundary conditions \( \lvert \uparrow \rangle \) and \( \lvert \downarrow \rangle \) in the Ising model constitutes a module category over fusion category that arises from the \( \mathbb{Z}_2 \) global symmetry \cite{Choi:2023xjw}. This relative entanglement sector plays a crucial role in understanding the quantum correlations present in the system, reflecting how the fixed boundary conditions influence the overall entanglement structure. The module category framework enables us to classify and analyze the different representations of the \( \mathbb{Z}_2 \) symmetry, highlighting the interplay between symmetry and entanglement in boundary conformal field theories.

It is well known that the low-energy physics of (chiral) gapped topologically ordered states can be described by a topological quantum field theory which possesses edge states described by a chiral (1+1)-dimensional conformal field theory
\cite{Witten:1988hf, Moore:1991ks}. The primary fields of the conformal field theory correspond to the quasiparticles present in the topological phase. In this context, a zero left-right relative entropy may indicate a fundamental connection between different phases of the theory, suggesting that the boundary states either belong to the same phase or are related through a symmetry transformation, providing insight into critical phenomena and phase transitions.

For instance, the Ising CFT describes the edge excitations of a \(p + ip\) superconductor \cite{Read:1999fn, Ivanov:2000mjr}. Different bulk excitations correspond to various conformal boundary conditions in the critical Ising model. Specifically, the absence of bulk quasiparticles or  fermionic
quasiparticles  corresponds to fixed boundary conditions,
%fixed up and fixed down boundary conditions,respectively. 
while free boundary conditions are associated with the presence of the vortex creation operator in the bulk \cite{Fendley:2009gm}. The relative entanglement sectors are related to a set of bulk quasiparticles which are grouped through special features.

Furthermore, the reduced density matrix of a finite spatial region within the gapped topological state corresponds to the thermal density matrix of the chiral edge state CFT located at the boundary of that region \cite{ludwig-qi}. This relationship invites further exploration into the connection between left-right relative entropy and the spatial relative entropy of (2+1)-dimensional topological quantum field theories \cite{Ghasemi:2022gh}.

Moreover, interpreting the left-right relative entropy for $D_{p}$-branes could be intriguing and provide a way to explore distinguishability measures in the context of string theory. It shall be interesting to probe the gravity dual of the LRRE using
the methods of AdS/CFT \cite{Takayanagi:2011zk}.
These issues will be addressed  in future works.

Our results suggest that chiral coarse-graining induces a nontrivial information-theoretic quotient of the space of boundary conditions. Understanding the resulting relative entanglement sectors in terms of generalized symmetries, topological defects, and categorical structures constitutes an interesting direction for future work.

\appendix
\label{App:}

\section{Positivity of the LRRE, Eq.(\ref{TRE-CARD})}
\label{App:Positivity}

Here, we prove the positivity of LRRE, Eq.(\ref{TRE-CARD}), as a measure of distinguishability.\\

\noindent\textbf{Proposition.} Let \(S\) be the modular \(S\)-matrix of a unitary diagonal CFT. For any labels \(a,b\),
\[
D_{\ket{\widetilde{h}_{l}},\ket{\widetilde{h}_{k}}}=\sum_j |S_{lj}|^2 \log\frac{|S_{lj}|^2}{|S_{kj}|^2} \ge 0,
\]
with equality iff \(|S_{lj}|^2 = |S_{kj}|^2\) for all \(j\).\\

\noindent\textbf{Proof.} By unitarity of \(S\) we have
\[
\sum_j S_{lj} S_{kj}^* = \delta_{lk}.
\]
Putting \(l=k\) gives \(\sum_j |S_{lj}|^2 = 1\) for each fixed \(l\). Define
\[
p_{l,j}:=|S_{lj}|^2,\qquad p_{k,j}:=|S_{kj}|^2.
\]
Then \(p_l\) and \(p_k\) are probability distributions (nonnegative and summing to \(1\)). The left-hand side is exactly the classical Kullback--Leibler divergence between \(p_l\) and \(p_k\):
\[
\sum_j |S_{lj}|^2 \log\frac{|S_{lj}|^2}{|S_{kj}|^2} = \sum_j p_{l,j}\log\frac{p_{l,j}}{p_{k,j}} = D_{\mathrm{KL}}(p_l\Vert p_k).
\]
The Kullback--Leibler divergence satisfies \(D_{\mathrm{KL}}(p_l\Vert p_k)\ge 0\), a result known as Gibbs' inequality, with equality iff \(p_l=p_k\). This proves the proposition.
\\

\section{Orthogonality of the Global States, infinity of Relative Entropy  vs vanishing of reduced Density Matrices}
\label{App:Orthogonality} 

\subsection{Setup with Bell States}

Consider the bipartite Hilbert space $\Hilbert{A}{B}$ with two different global pure states:

\begin{align}
	|\psi_1\rangle &= |\Phi^+\rangle = \frac{1}{\sqrt{2}}(|00\rangle + |11\rangle) \\
	|\psi_2\rangle &= |\Phi^-\rangle = \frac{1}{\sqrt{2}}(|00\rangle - |11\rangle)
\end{align}

\subsection{Global Analysis}

The global density matrices are:
\begin{align}
	\rho_{\text{global}}^1 &= |\Phi^+\rangle\langle\Phi^+| \\
	\rho_{\text{global}}^2 &= |\Phi^-\rangle\langle\Phi^-|
\end{align}

These states are \textbf{orthogonal}:
\begin{equation}
	\langle\Phi^+|\Phi^-\rangle = \frac{1}{2}(\langle 00| - \langle 11|)(|00\rangle + |11\rangle) = 0
\end{equation}

Therefore, their global relative entropy is infinite:
\begin{equation}
	\relent{\rho_{\text{global}}^1}{\rho_{\text{global}}^2} = +\infty
	\label{eq:global_infinite}
\end{equation}

\subsection{Local Analysis via Partial Trace}

The reduced density matrices on subsystem A are:
\begin{align}
	\rho_A^1 &= \operatorname{Tr}_B(|\Phi^+\rangle\langle\Phi^+|) = \frac{1}{2}(|0\rangle\langle 0| + |1\rangle\langle 1|) = \frac{I_A}{2} \\
	\rho_A^2 &= \operatorname{Tr}_B(|\Phi^-\rangle\langle\Phi^-|) = \frac{1}{2}(|0\rangle\langle 0| + |1\rangle\langle 1|) = \frac{I_A}{2}
\end{align}
The coefficients of the expansion becomes same for both states after taking partial trace.
Since $\rho_A^1 = \rho_A^2$, their relative entropy is zero:
\begin{equation}
	\relent{\rho_A^1}{\rho_A^2} = 0.
	\label{eq:local_zero}
\end{equation}

The results (\ref{eq:global_infinite}) and (\ref{eq:local_zero}) are mathematically consistent due to:

\textbf{	Monotonicity under Partial Trace}:
For any quantum channel $\mathcal{E}$ (including partial trace):
\begin{equation}
	\relent{\mathcal{E}(\rho)}{\mathcal{E}(\sigma)} \leq \relent{\rho}{\sigma}
\end{equation}
Specifically:
\begin{equation}
	\relent{\rho_A^1}{\rho_A^2} \leq \relent{\rho_{\text{global}}^1}{\rho_{\text{global}}^2} \quad \Rightarrow \quad 0 \leq +\infty
\end{equation}

\subsection{Physical Interpretation}

The paradox arises from confusing different levels of description:

\begin{itemize}
	\item \textbf{Local Observer on A:} Only has access to $\rho_A = I_A/2$. All measurements yield identical statistics for both global preparations.
	\item \textbf{Global Observer:} Has access to correlations between A and B and can perform Bell measurements to perfectly distinguish the states.
\end{itemize}

\acknowledgments

It is my pleasure to thank Pasquale Calabrese,  Sepideh Mohammadi, Ali Mollabashi, Ahmad Moradpouri, Sara Murciano, and Abolhassan Vaezi for  reading our manuscript and providing insightful comments. We are sincerely grateful to 
Horacio Casini, Sepideh Mohammadi, and Edward Witten for their valuable comments and insightful suggestions.

\end{document}